\documentclass[preprint,pra,amsfonts,amssymb,10pt,showpacs,showkeys,byrevtex,onecolumn,superscriptaddress]{revtex4-1}
\usepackage{amsmath}
\usepackage{graphicx}
\usepackage{amsfonts}
\usepackage{amssymb}%
\pdfoutput=1 
\providecommand{\U}[1]{\protect\rule{.1in}{.1in}}

\begin{document}
\title{Spectral properties of correlation functions of fields with arbitrary position
dependence in restricted geometries from the ballistic to the diffusive regimes.}
\author{C.M. Swank}
\affiliation{Physics Department, North Carolina State University, Raleigh, NC 27695}
\author{A. Petukhov}
\affiliation{Institut Laue-Langevin, BP156, 38042 Grenoble Cedex 9, France }
\author{R. Golub}
\affiliation{Physics Department, North Carolina State University, Raleigh, NC 27695}

\begin{abstract}
The transition between ballistic and diffusive motion poses difficult problems
in several fields of physics. In this work we show how to calculate the
spectra of the correlation functions between fields of arbitrary spatial
dependence as seen by particles moving through the fields  in regions bounded by specularly
reflecting walls valid for diffusive and ballistic motion as well as the
transition region in between for motiion in 2 and 3 dimensions.

Applications to relaxation in nmr are discussed.

\end{abstract}

\maketitle

\section{\bigskip Introduction}

The problem of the transition between ballistic and diffusive motion impacts
many fields of physics including charge transport in semiconductors, electron
transport in mesoscopic systems, fluid flow, heat conduction \cite{Maris},
light transport in random media, \cite{Elal}, \cite{Durian} relaxation and
frequency shifts in magnetic resonance of particles moving in non-uniform
fields \cite{Cates}, \cite{Relax}. In this work we concentrate on the latter
problem and present solutions in finite regions in 1, 2 and 3 dimensions valid
for ballistic and diffusive \ motions and the crossover between.

In these problems, the physical effects of interest depend on the spectra of
the correlation functions of various field components as seen by the moving
particles.%
\begin{equation}
S_{1,2}\left(  \omega\right)  =\int_{-\infty}^{\infty}\left\langle
B_{1}\left(  t\right)  B_{2}\left(  t+\tau\right)  \right\rangle
e^{-i\omega\tau}d\tau \label{2}%
\end{equation}
where $\left\langle {}\right\rangle $ signifies the ensemble average over the
gas particles.

As shown first by \cite{Wayne} (see also \cite{Tarc}, \cite{Bloom}) and then
applied by \cite{Mcgreg}, \cite{Petuk} and \cite{Clayt} \ these correlation
functions are given by the conditional probability $p\left(  \overrightarrow
{r},\tau|\overrightarrow{r}_{0},0\right)  $, the probability that a particle
at $\overrightarrow{r}_{0}$ at $t=0$ will be found at a position
$\overrightarrow{r}$ at the later time $\tau$:%
\begin{equation}
R_{1,2}\left(  \tau\right)=
\left\langle B_{1}\left(  t\right)  B_{2}\left(  t+\tau\right)  \right\rangle
=\int\int d^{3}rd^{3}r_{0}B_{1}\left(  \overrightarrow{r}_{0}\right)
B_{2}\left(  \overrightarrow{r}\right)  p\left(  \overrightarrow{r},t_{0}%
+\tau|\overrightarrow{r}_{0},t_{0}\right)  p\left(  \overrightarrow{r}%
_{0},t_{0}\right)  \label{1}%
\end{equation}
where $p\left(  \overrightarrow{r}_{0},t_{0}\right)  $ is the probability of
finding a particle at $\overrightarrow{r}_{0}$ at time $t_{0}$ which is taken
as constant $\left(  =1/L\text{ in 1 dimension}\right)  $. In this paper we
consider single-speed transport only.

We treat the motion as a persistent continuous time random walk with isotropic
scattering (in 2 and 3 dimensions) and uniform velocity with an exponential
distribution of times, $t$, between scatterings,%

\begin{equation}
\psi\left(  t\right)  =\frac{1}{\tau_{c}}e^{-t/\tau_{c}} \label{3}%
\end{equation}

\section{Motion in 1 dimension}

Goldstein \cite{Goldst} has shown that in these conditions the conditional
probability distribution, considering the motion as the continuum limit of a
persistent random walk, satisfies the telegrapher's equation in one dimension.
As shown by Masoliver et al \cite{MasPoWei} the equation can be solved by
separation of variables and the conditional probability with reflecting
boundary conditions at $x=0,L$ is given by
\begin{equation}
p(x,t~|~x_{0},0)=\frac{1}{L}\left\{  \sum_{n}\cos\left(  \frac{n\pi x}%
{L}\right)  \cos\left(  \frac{n\pi x_{0}}{L}\right)  \right\}  \left\{
\cosh\left(  \frac{s_{n}t}{2\tau_{c}}\right)  +\frac{1}{s_{n}}\sinh\left(
\frac{s_{n}t}{2\tau_{c}}\right)  \right\}  e^{-\frac{t}{2\tau_{c}}} \label{5}%
\end{equation}
with $s_{n}=\sqrt{1-4\omega_{n}^{2}\tau_{c}^{2}}$ and $\omega_{n}=n\pi v/L$
with $v$ the velocity of the particles. \ Note the change in behavior as
$s_{n}$ goes from real to imaginary. The behavior of this solution is
discussed in some detail in \cite{MasPoWei} and \cite{GSxxx}. There is a delta
function peak leaving the source with velocity, $v,$ This peak diminishes in
time as particles are scattered and a wake of scattered particles builds up
behind the peak. The motion can be followed through successive wall reflections.

Then according to (\ref{1}) the field correlation function is given by%
\begin{equation}
R_{1,2}\left(  \tau\right)
=\sum_{n}F_{1}\left(  k_{n}\right)  F_{2}\left(  k_{n}\right)  \left\{
\cosh\left(  \frac{s_{n}\tau}{2\tau_{c}}\right)  +\frac{1}{s_{n}}\sinh\left(
\frac{s_{n}\tau}{2\tau_{c}}\right)  \right\}  e^{-\frac{\tau}{2\tau_{c}}}%
\end{equation}

with 
\begin{equation}
F_{1,2}\left(  k_{n}\right)  =\frac{1}{L}\int_{0}^{L}dxB_{1,2}\left(
x\right)  \cos\left(  k_{n}x\right)%  
\end{equation}

 and
 \begin{equation}
  k_{n}=n\pi/L%
 \end{equation}

   The Fourier
transform of this gives the desired spectrum which determines physical
phenomena.%
\begin{equation}
S_{1,2}\left(  \omega\right)  =\sum_{n}F_{1}\left(  k_{n}\right)  F_{2}\left(
k_{n}\right)  \left(  \frac{2\omega_{n}^{2}\tau_{c}}{\omega^{2}\allowbreak
+\left[  \left(  \omega^{2}-\omega_{n}^{2}\right)  \right]  ^{2}\tau_{c}^{2}%
}\right)
\end{equation}

\bigskip For not too high $\omega$: $\omega\ll\pi v/L$ this result is
consistent with one obtained earlier for the diffusion regime of motion
\cite{Petuk}, \cite{Clayt}.

This result is consistent with calculations of the position- correlation
function, $B_{1}=B_{2}\sim x$ for various values of normalized mean free path,
$l^{\prime}=v\tau_{c}/L$ presented in \cite{Relax}, see figure 2 in that
paper. Notice as the motions approaches ballistic, $l^{\prime}>>1,$ the
relaxation shows a resonance behavior \ when a harmonic of the wall collision
frequency coincides with the Larmor frequency, which has not been noted
previously. The resonance peaks are smoothed by averaging over the velocity
distribution. Experiments to observe this are under way.

\bigskip

It may be shown that for a perturbation field with a uniform gradient
$B=Gx$\ the series (6) can be summed in closed form:%
\begin{equation}
S=2\left(  \frac{G}{\omega}\right)  ^{2}\frac{v^{2}\tau_{c}}{\left(
1+\omega\tau_{c}\right)  ^{2}}F\left(  \omega\right)  \label{9}%
\end{equation}
where%
\begin{align}
F\left(  \omega\right)   &  =1-\operatorname{Im}\left[  \left(  \omega\tau
_{c}+i\right)  \frac{\tan\left(  \frac{\omega\tau_{b}}{2}\sqrt{1-i/\omega
\tau_{c}}\right)  }{\left(  \frac{\omega\tau_{b}}{2}\sqrt{1-i/\omega\tau_{c}%
}\right)  }\right] \\
\tau_{b}  &  =\frac{L}{v}\quad characteristic~ballistic~time\nonumber
\end{align}
In the limit $\omega\tau_{c}\ll1$ (\ref{9}) goes over to the diffusion theory
result, see \cite{Mcgreg}.

\section{Motion in 2 dimensions.}

\subsection{Solution in free space}

The obvious extension of the above ideas is to apply them to higher dimensions
by writing the Telegrapher's equation (TE) with the appropriate form of
$\triangledown^{2}$. This was first suggested by Mark Kac, \cite{Kac}. However
the solution of the TE cannot represent a conditional probability in two or
higher dimensions as it can take on negative values, \cite{PorrMasWeiss} .
Morse and Feshbach \cite{M and F} discuss the interesting properties of
solutions of the TE (as well the wave equation) in 2 dimensions. This point
seems to have been missed in an otherwise interesting and useful work
\cite{KP}, which nonetheless finds the same solution for the spectrum of the
conditional probability function as given by \cite{MasPoWei}. In a remarkable
paper Masoliver et al \cite{MasPoWei2} have shown that the conditional
probability function for a persistent random walk in two dimensions, with the
time between scattering distributed according to (\ref{3}) and a uniform
distribution of scattering angles (s-wave scattering) for particles starting
at the origin of coordinates with velocity $v$, satisfies the 2 D TE with an
additional source term%
\begin{equation}
\rho\left(  r,t\right)  =\frac{v^{2}}{2\pi r}e^{-t/\tau_{c}}\frac{\partial
}{\partial r}\left(  \frac{\delta\left(  r-vt\right)  }{r}\right)
\end{equation}
The source moves along with the unscattered particles and represents the
scattered particles that are not accounted for in the homogeneous TE. The
authors give the solution for motion in the infinite domain as:
\begin{equation}
P_{02}\left(  r,t\right)  =e^{-t/\tau_{c}}\left[  \frac{\delta\left(
r-vt\right)  }{2\pi r}+\frac{1}{2\pi v\tau_{c}\sqrt{\left(  vt\right)
^{2}-r^{2}}}e^{\frac{\sqrt{\left(  vt\right)  ^{2}-r^{2}}}{v\tau_{c}}}%
\Theta\left(  vt-r\right)  \right]
\end{equation}
where $\Theta$ is the unit step function. The authors also give the
Fourier-Laplace transform of the solution
\begin{align}
\hat{P}_{02}\left(  \overrightarrow{Q},s=i\omega\right)   &  =\int d^{2}%
r\int_{0}^{\infty}p\left(  r,t\right)  e^{\left(  i\overrightarrow{Q}%
\cdot\overrightarrow{r}-st\right)  }dt\nonumber\\
&  =\frac{\tau_{c}}{\left[  \left(  1+i\omega\tau_{c}\right)  ^{2}+v^{2}%
\tau_{c}^{2}\left\vert \overrightarrow{Q}\right\vert ^{2}\right]  ^{1/2}-1}
\label{8}%
\end{align}
The expression (\ref{8}) and the equivalent one for three dimensions have been
rederived by Kolesnik, \cite{Kolesnik}, who gives a general treatment
applicable to an arbitrary number of dimensions.

\subsection{Solution in bounded rectangular region.}

We now extend these results to find the spectrum of the field correlation function in a rectangular region
bounded by reflecting walls. This solution is presented here for the first
time. As we desire the spectrum of a field correlation function, we will work
with the Fourier transform of $P_{02}\left(  \overrightarrow{x},\tau\right)
,$ where we are using the $0$ subscript to denote the free space solution,%
\begin{equation}
P_{02}\left(  \overrightarrow{x},\tau\right)  =\frac{1}{\left(  2\pi\right)
^{3}}\int d^{2}Q\int d\omega\widetilde{P}_{0}\left(  \overrightarrow{Q}%
,\omega\right)  e^{-i\left(  \overrightarrow{Q}\cdot\overrightarrow{x}%
-\omega\tau\right)  } \label{6}%
\end{equation}
where $\widetilde{P}_{02}\left(  \overrightarrow{Q},\omega\right)  =\hat
{P}_{02}\left(  \overrightarrow{Q},s=i\omega\right)  $ (\ref{8})

We use the method of images to find the conditional probability function in the presence of the
rectangular boundaries at $x=\pm L_{x}/2,$ $y=\pm L_{y}/2.$ The location of
the images is sketched in fig.1. Wall reflections are taken into account by considering particles coming from the image sources. The desired conditonal probability function for a restrcited rectangular domain (with specularly reflecting walls) is given by the superposition of probability coming from the original source and all the images.

%TCIMACRO{\FRAME{ftbpFU}{5.8281in}{4.4765in}{0pt}{\Qcb{A) Showing the physical
%region (shaded) and a few of the periodic repetition cells. A source point and
%its images are shown. We consider the conditional probability function as a
%wave travelling from the source to the obsrvation point, P. It is useful to
%use reciprocity and consider the wave as travelling from P to the source. The
%solution is then given by considering the wave as travelling in an
%unrestricted domain and taking the sum of all the waves reaching all the image
%points. In travelling from the physical source \ to the wall at x=L/2 in the
%physical case with boundaries, the particles see x increasing to the right.
%Trajectories in this region are replaced by trajectories in the region between
%the image point and the wall so in this region the effective value of x=x must
%be seen to be increasing going from the image to the wall. Thus x has to be
%taken as periodic with period 2L. An arbitrary field depending on position
%must be treated in the same way, ( as shown in B) which is a plot of an
%arbitrary field varying with x vs. x (solid line)). and hence must be taken as
%periodic.}}{}{fig1.jpg}{\special{ language "Scientific Word";
%type "GRAPHIC";  maintain-aspect-ratio TRUE;  display "USEDEF";
%valid_file "F";  width 5.8281in;  height 4.4765in;  depth 0pt;
%original-width 10.0124in;  original-height 7.6803in;  cropleft "0";
%croptop "1";  cropright "1";  cropbottom "0";
%filename 'fig1.jpg';file-properties "XNPEU";}}}%
%BeginExpansion
\begin{figure}
[ptb]
\begin{center}
\includegraphics[
width=\textwidth
]%
{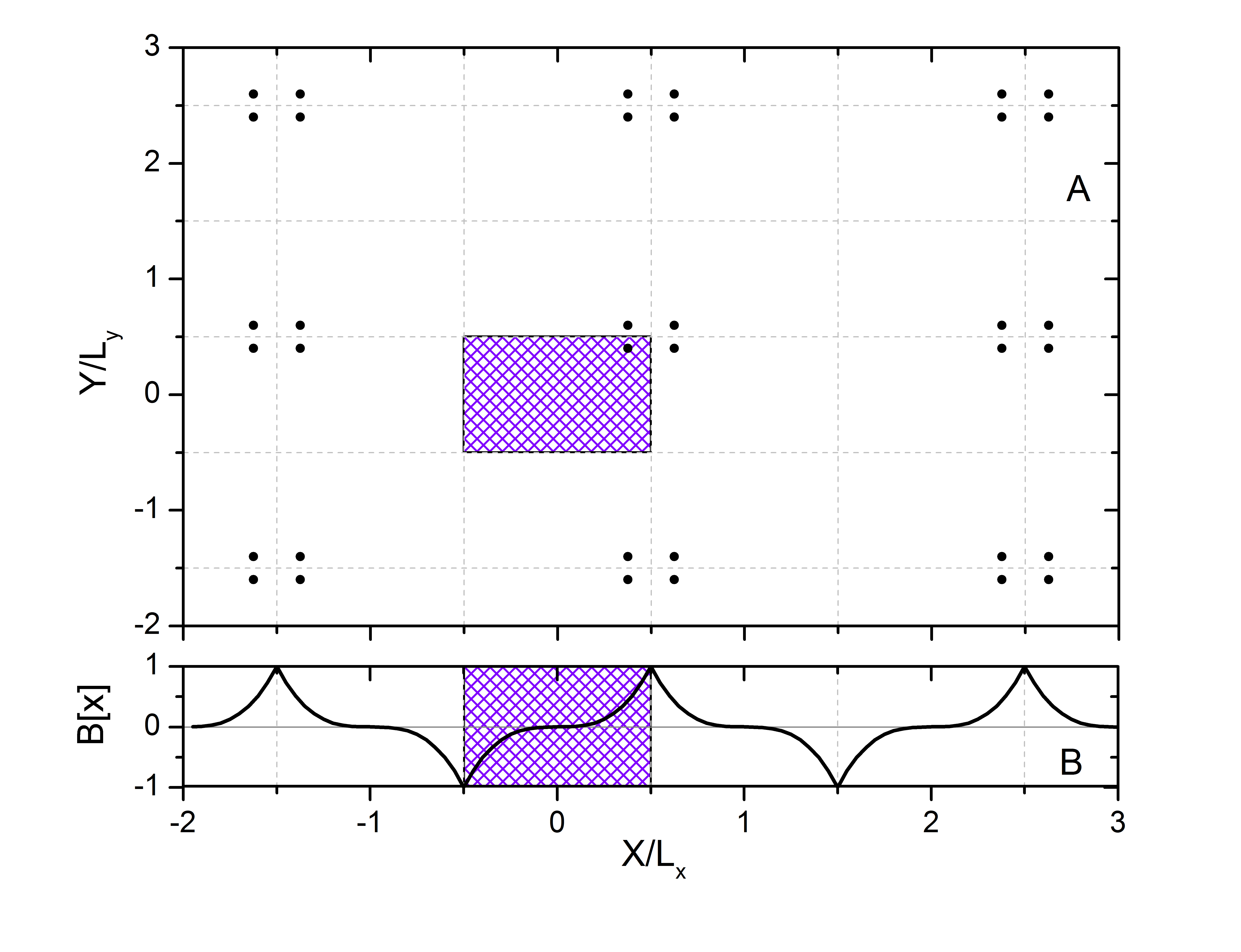}%
\caption{A) Showing the physical region (shaded) and a few of the periodic
repetition cells. A source point and its images are shown. We consider the
conditional probability function as a wave travelling from the source to the
obsrvation point, P. It is useful to use reciprocity and consider the wave as
travelling from P to the source. The solution is then given by considering the
wave as travelling in an unrestricted domain and taking the sum of all the
waves reaching all the image points. In travelling from the physical source
\ to the wall at x=L/2 in the physical case with boundaries, the particles see
x increasing to the right. Trajectories in this region are replaced by
trajectories in the region between the image point and the wall so in this
region the effective value of x=x must be seen to be increasing going from the
image to the wall. Thus x has to be taken as periodic with period 2L. An
arbitrary field depending on position must be treated in the same way, ( as
shown in B) which is a plot of an arbitrary field varying with x vs. x (solid
line)). and hence must be taken as periodic.}%
\end{center}
\end{figure}
%EndExpansion

We see that the physical region is repeated periodically throughout the plane
and there is one image point in each cell (the positions of the images are not
the same in every cell). \ Then we can write the solution as
\begin{equation}
P\left(  \overrightarrow{r},\tau|\overrightarrow{r_{0}},0\right)  =\sum_{i}P_{02}\left(
\overrightarrow{r}-\overrightarrow{r}_{0,i},\tau\right)  \label{10}%
\end{equation}
where $\overrightarrow{r}_{0,i}$ is the location of the $i^{th}$ image. ($i=0$ denotes the physical source).

Fig. 2A shows the behavior of $P_{02}\left(  \overrightarrow{r},\tau\right)  ,
$ (\ref{10}) for the parameters, $v=1,$ $L=2,$ $\lambda=0.2$ and $t=2.7t_{b}
$, where $t_{b}=L/2v.$%
Figure 2B show a cross section along the x axis of the function shown in fig.2A.

\bigskip Then (\ref{1})
\begin{equation}
R_{1,2}\left(  \tau\right)=\left\langle B_{1}\left(  t\right)  B_{2}\left(  t+\tau\right)  \right\rangle
=\int\int d^{2}rd^{2}r_{0}B_{1}\left(  \overrightarrow{r}_{0}\right)
B_{2}\left(  \overrightarrow{r}\right)  \sum_{i}P_{02}\left(  \overrightarrow
{r}-\overrightarrow{r}_{0,i},\tau\right)  p\left(  \overrightarrow{r}%
_{0,i},t_{0}\right)  \label{4}%
\end{equation}
where the integrals are taken over the physical cell, $\left(  p\left(
\overrightarrow{r}_{0,i},t_{0}\right)  =\frac{1}{L_{x}L_{y}}\right)  .$

Integrating over $d^{2}r_{0}$ in the physical cell means that each image point
in each cell will cover its entire cell and the sum over images and
integration over the physical cell can be calculated by taking the function
$P_{02}$ to be the infinite domain function integrated over all space if we
continue the field periodically as explained in fig.1. This idea was
introduced by Wayne and Cotts \cite{Wayne} (see also \cite{Tarc})

We can rewrite (\ref{4}) as%
\begin{equation}
R_{1,2}\left(  \tau\right)
=\frac{1}{L_{x}L_{y}}\int_{-L_{x}/2}^{L_{x}/2}\int_{-L_{y}/2}^{L_{y}%
/2}dxdy\int_{-\infty}^{\infty}\int_{-\infty}^{\infty}dx_{0}dy_{0}\widetilde
{B}_{1}\left(  \overrightarrow{r}_{0}\right)  B_{2}\left(  \overrightarrow
{r}\right)  P_{02}\left(  \overrightarrow{r}-\overrightarrow{r}_{0}%
,\tau\right)
\end{equation}
where $\widetilde{B}_{1}\left(  \overrightarrow{r}_{0}\right)  $ is the
periodic extension of $B_{1}\left(  \overrightarrow{r}_{0}\right)  $ beyond
the physical cell.

Writing $P_{02}\left(  \overrightarrow{r}-\overrightarrow{r}_{0},\tau\right)
$ in terms of its Fourier transform (\ref{6}) we find for the spectrum of the
correlation function ($l_{x},l_{y}$ are integers)%
\begin{equation}
S_{1,2}\left(  \omega\right)  =\frac{1}{L_{x}L_{y}}\sum_{l_{x},l_{y}}%
\beta_{l_{x},l_{y}}B_{2}\left(  \overrightarrow{q}_{l_{x},l_{y}}\right)
\widetilde{P}_{02}\left(  \overrightarrow{q}_{l_{x},l_{y}},\omega\right)
\label{11}%
\end{equation}
with $\overrightarrow{q}_{l_{x},l_{y}}=\left(  \frac{l_{x}\pi}{L_{x}}%
,\frac{l_{y}\pi}{L_{y}}\right)  ,$ $\beta_{l_{x},l_{y}}=\frac{1}{L_{x}L_{y}%
}\int_{-L_{x}/2}^{3L_{x}/2}\int_{-L_{y}/2}^{3L_{y}/2}dx_{0}dy_{0}%
e^{i\overrightarrow{q}_{l_{x},l_{y}}\cdot\overrightarrow{x}_{o}}B_{1}\left(
\overrightarrow{x}_{o}\right)  $ where the integral is taken over one period and

$B_{2}\left(  \overrightarrow{q}_{l_{x},l_{y}}\right)  ==\int_{-L_{x}%
/2}^{L_{x}/2}\int_{-L_{y}/2}^{L_{y}/2}dxdyB_{2}\left(  \overrightarrow
{x}\right)  e^{-i\left(  \overrightarrow{q}_{l_{x},l_{y}}\cdot\overrightarrow
{x}\right)  }.$

%TCIMACRO{\FRAME{ftbpFU}{3.364in}{4.4765in}{0pt}{\Qcb{A) Showing a snapshot of
%\ $P\left(  \overrightarrow{r},\tau\right)  $ for parameters given in the
%text. The white lines are the peak of unscattered particles while the brightness
%of the contrast indicates the build up of the wake of scattered particles.
%\ \ B)\ \ A cross section along the $x$ axis of the function $P_{02}\left(
%\overrightarrow{r},\tau=2.7\right)  $ shown in A). The smaller peaks at
%$\pm0.7$ are \ the particles that made a collison with the walls at $x=\pm1.$
%The larger peaks represent the crossing of the two "waves" emanating from the
%image points at $\left(  \pm2,\pm2\right)  $ and are larger because they are
%the sum of the two "waves". Solid lines are the result of equation (\ref{10})
%while the dots are the results of a Monte-Carlo simulation.}}{}{graph2.jpg}%
%{\special{ language "Scientific Word";  type "GRAPHIC";
%maintain-aspect-ratio TRUE;  display "USEDEF";  valid_file "F";
%width 3.364in;  height 4.4765in;  depth 0pt;  original-width 6.4798in;
%original-height 8.6401in;  cropleft "0";  croptop "1";  cropright "1";
%cropbottom "0";  filename 'Graph2.jpg';file-properties "XNPEU";}}}%
%BeginExpansion
\begin{figure}
[ptb]
\begin{center}
\includegraphics[scale=0.8]{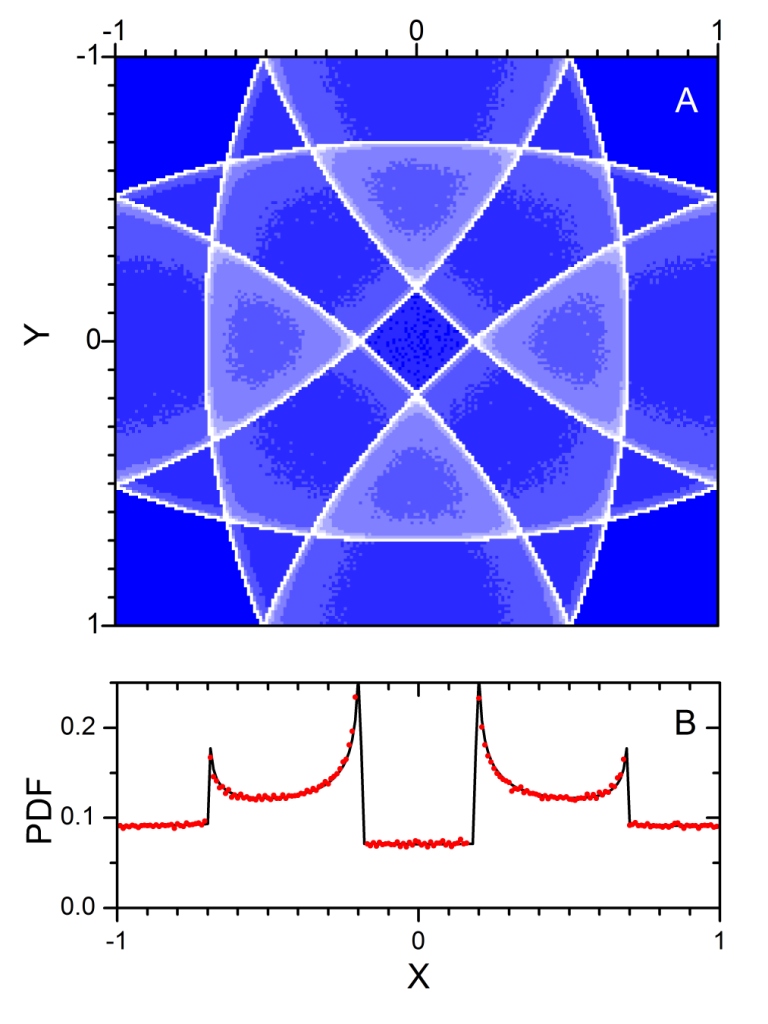}%
\caption{A) Showing a snapshot of \ $P\left(  \protect \overrightarrow{r},\tau\right)
$ for parameters given in the text. The white lines are the peak of
unscattered particles while the brightness of the contrast indicates the build
up of the wake of scattered particles. \ \ B)\ \ A cross section along the $x$
axis of the function $P_{02}\left(  \protect\overrightarrow{r},\tau=2.7\right)  $
shown in A). The smaller peaks at $\pm0.7$ are \ the particles that made a
collison with the walls at $x=\pm1.$ The larger peaks represent the crossing
of the two "waves" emanating from the image points at $\left(  \pm
2,\pm2\right)  $ and are larger because they are the sum of the two "waves".
Solid lines are the result of equation (\ref{10}) while the dots are the
results of a Monte-Carlo simulation.}%
\end{center}
\end{figure}
%EndExpansion

\section{Solution in 3 dimensions}

\subsection{Solution in free space}

Masoliver et al \cite{MasPoWei2} (see also \cite{ClaesVan}) give the form of
the Fourier Laplace transform of the conditional probability function in three
dimensions as
\begin{equation}
\hat{P}_{03}\left(  \overrightarrow{Q},s\right)  =\int d^{3}r\int_{0}^{\infty
}p\left(  r,t\right)  e^{\left(  i\overrightarrow{Q}\cdot\overrightarrow
{r}-st\right)  }dt=\tau_{c}\frac{\tan^{-1}\left(  \frac{Qv\tau_{c}}%
{1+s\tau_{c}}\right)  }{\left(  Qv\tau_{c}-\tan^{-1}\left(  \frac{Qv\tau_{c}%
}{1+s\tau_{c}}\right)  \right)  }%
\end{equation}
The same result was later obtained by \cite{Kolesnik} using a different method
valid for any number of dimensions.

\subsection{Solution in a rectangular box}

The same reasoning applies here as in 2 dimensions above, the crucial point
being that there is one image point per cell.

The spectrum of the correlation function is then:%
\begin{equation}
S_{1,2}\left(  \omega\right)  =\sum_{l_{x},l_{y},l_{z}}\beta_{l_{x}%
,l_{y},l_{z}}B_{2}\left(  \overrightarrow{q}_{l_{x},l_{y},l_{z}}\right)
\widetilde{P}_{03}\left(  \overrightarrow{q}_{l_{x},l_{y},l_{z}}%
,s=i\omega\right)
\end{equation}

where $\overrightarrow{q}_{l_{x},l_{y},l_{z}}=\left(  \frac{l_{x}\pi}{L_{x}%
},\frac{l_{y}\pi}{L_{y}},\frac{l_{z}\pi}{L_{z}}\right)  ,$
\begin{equation}
\beta_{l_{x},l_{y},l_{z}}=\frac{1}{V}\int_{-L_{x}/2}^{3L_{x}/2}\int_{-L_{y}%
/2}^{3L_{y}/2}\int_{-L_{z}/2}^{3L_{z}/2}dx_{0}dy_{0}dz_{0}e^{i\overrightarrow
{q}_{l_{x},l_{y},l_{z}}\cdot\overrightarrow{x}_{o}}\widetilde{B}_{1}\left(
\overrightarrow{x}_{o}\right)
\end{equation}
(each integration is over one complete period) and%

\begin{equation}
B_{2}\left(  \overrightarrow{q}_{l_{x},l_{y},l_{z}}\right)  =\frac{1}{V}%
\int_{-L_{x}/2}^{L_{x}/2}\int_{-L_{y}/2}^{L_{y}/2}\int_{-L_{z}/2}^{L_{z}%
/2}dxdydzB_{2}\left(  \overrightarrow{x}\right)  e^{-i\left(  \overrightarrow
{q}_{l_{x},l_{y},l_{z}}\cdot\overrightarrow{x}\right)  }%
\end{equation}

\section{\bigskip Applications}

In the following we show some illustrative examples of the use of the above technique.

\subsection{Position auto-correlation functions (Uniform gradient field)}

In Fig. 3A) we show the calculation for the spectrum of the $x-x$
auto-correlation function, determining relaxation times for a field with
uniform gradient, as is usually assumed in studies of relaxation, for the one
dimensional case for various values of damping using (\ref{9}).%

%TCIMACRO{\FRAME{ftbpFU}{5.8788in}{6.9099in}{0pt}{\Qcb{Showing the spectum of
%the autocorrelation function of position $x$ for various values of the
%collison time, $\tau_{c}.$ vs. dimensionless frequency $\omega\tau_{b.}$
%\ \ \ \ \ \ \ \ \ \ \ \ \ \ \ \ \ \ A) For one dimension, B )The same for a
%square in 2 dimensions, \ \ C)For a cube in three
%dimensions\ \ \ \ \ \ \ \ \ \ \ }}{}{fig3.jpg}%
%{\special{ language "Scientific Word";  type "GRAPHIC";  display "USEDEF";
%valid_file "F";  width 5.8788in;  height 6.9099in;  depth 0pt;
%original-width 7.6803in;  original-height 10.6849in;  cropleft "0";
%croptop "1";  cropright "1";  cropbottom "0";
%filename 'fig3.jpg';file-properties "XNPEU";}}}%
%BeginExpansion
\begin{figure}
[ptb]
\begin{center}
\includegraphics[scale=0.8]{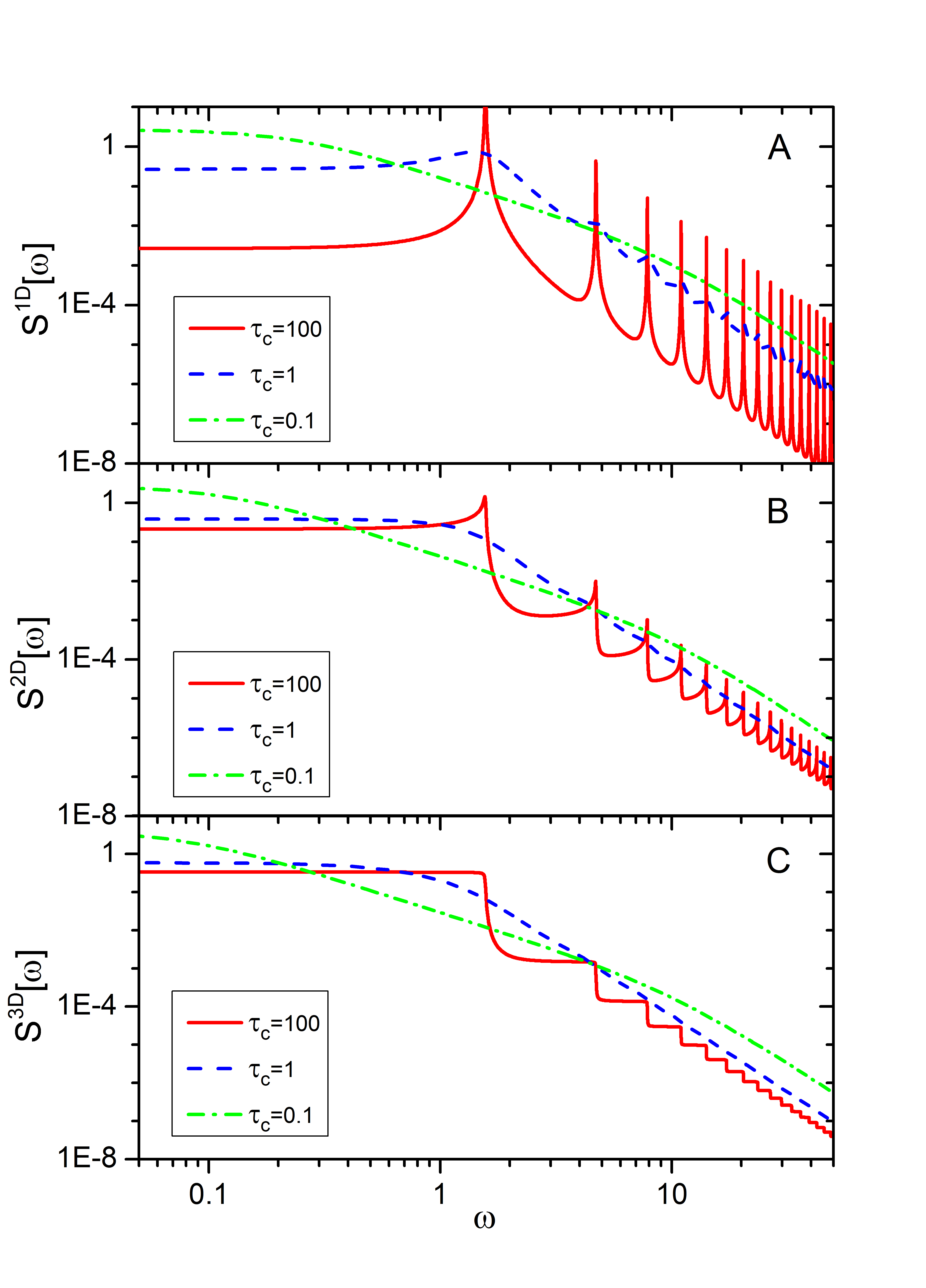}%
\caption{Showing the spectum of the autocorrelation function of position $x$
for various values of the collison time, $\tau_{c}.$ vs. dimensionless
frequency $\omega\tau_{b.}$ \ \ \ \ \ \ \ \ \ \ \ \ \ \ \ \ \ \ A) For one
dimension, B )The same for a square in 2 dimensions, \ \ C)For a cube in three
dimensions\ \ \ \ \ \ \ \ \ \ \ }%
\end{center}
\end{figure}
%EndExpansion

The results for 2 and 3 dimensions are:

2 dimensions:%
\begin{equation}
S_{xx}^{\left(  2D\right)  }\left(  \omega\right)  =\frac{8}{\pi^{4}}L_{x}%
^{2}\sum_{n=odd}\frac{1}{n^{4}}\operatorname{Re}\left[  \hat{P}_{02}\left(
q_{n},s=i\omega\right)  \right]
\end{equation}

3 dimensions:%
\begin{align}
S_{xx}^{\left(  3D\right)  }\left(  \omega\right)   &  =\frac{8}{\pi^{4}}%
L_{x}^{2}\sum_{n=odd}\frac{1}{n^{4}}\operatorname{Re}\left[ \hat{P}_{03}\left(  \overrightarrow{q_{n}},s=i\omega\right)   \right] \\
\overrightarrow{q_{n}}  &  =\left[ \frac{n_{x}\pi}{L_{x}},\frac{n_{y}\pi}{L_{y}},\frac{n_{z}\pi}{L_{z}} \right]%
\end{align}

For large $\tau_{c}$ $\left(  \lambda\gg L\right)  $, we see that there is a
series of resonances, which will be broadened by averaging over a realistic
velocity distribution, and non-specular wall reflections. These are Rabi
resonances which occur when a harmonic of the periodic motion coincides with
the Larmor frequency. Similar resonances have been noted in cylindrical
geometry in another context \cite{JMP}. For $\tau_{c}=1$ the resonances are
still slightly visible, while at $\tau_{c}=0.1$ we have approached the
diffusion behavior. In 2 dimensions the resonances are less peaked because
closed orbits are less probable. In 3D the peaks (even in the undamped case)
are reduced to steps.

In the diffusion limit (green curves) we see the usual behavior at low
frequencies approaching a constant as $\omega\rightarrow0$ (non-adiabatic
region)$.$ For higher frequencies the spectrum starts to decrease as $\left(
1/\omega^{2}\right)  $, (adiabatic regime) and at still higher frequencies
$\left(  \omega\tau_{c}\gg1\right)  $ it falls as $\left(  1/\omega
^{4}\right)  $, the super-adiabatic region \cite{ScherWal}. For moderate
(blue) and light (red) damping, the $1/\omega^{2}$ behavior disappears and we
go directly from a constant to $1/\omega^{4}$ behavior.

\subsection{Short range interaction between the walls and the spin of the
particles}

This problem has been discussed in \cite{Petuk} using the conditional
probability density from diffusion theory. Here in figure 4 we present the
results valid for the diffusion and ballistic cases and the transition in
between calculated using the 2 dimensional conditional probability (\ref{11})
taking the interaction to be represented by a effective magnetic field:%

\begin{equation}
B_{eff}\left(  x\right)  =b_{a}\left(  e^{-\left(  L/2+x\right)  /\lambda
}-e^{-\left(  L/2-x\right)  /\lambda}\right)  \label{12}%
\end{equation}

where the range of the interaction is $\lambda.$%

%TCIMACRO{\FRAME{ftbpFU}{5.8281in}{4.4765in}{0pt}{\Qcb{Results of the
%auto-correlation function of the field given in equation (\ref{12}) for the
%range of force, $\lambda=.005,$ normalized to the size of the box and various
%valus of $\tau_{c}.$}}{}{fig4.jpg}{\special{ language "Scientific Word";
%type "GRAPHIC";  maintain-aspect-ratio TRUE;  display "USEDEF";
%valid_file "F";  width 5.8281in;  height 4.4765in;  depth 0pt;
%original-width 10.0124in;  original-height 7.6803in;  cropleft "0";
%croptop "1";  cropright "1";  cropbottom "0";
%filename 'Fig4.jpg';file-properties "XNPEU";}}}%
%BeginExpansion
\begin{figure}
[ptb]
\begin{center}
\includegraphics[
width=\textwidth
]%
{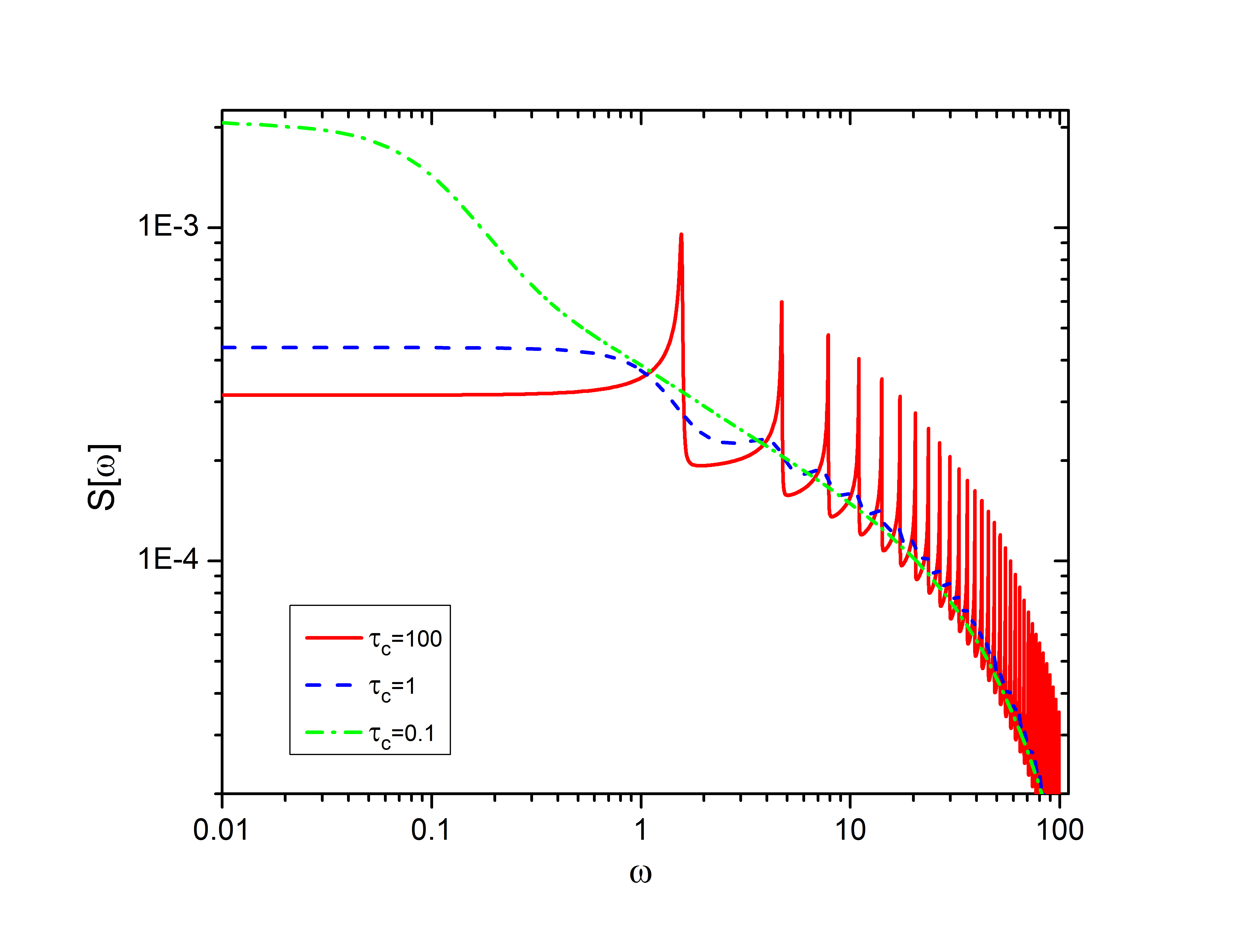}%
\caption{Results of the auto-correlation function of the field given in
equation (\protect\ref{12}) for the range of force, $\lambda=.005,$ normalized to the
size of the box and various valus of $\tau_{c}.$}%
\end{center}
\end{figure}
%EndExpansion

Here we see the influence of the shape of the field on the spectrum.

In the medium and underdamped cases the constant behavior goes over, as frequency increases, to a
$\left(  1/\sqrt{\omega}\right)  $ behavior before reaching the
super-adiabatic region $\left(  1/\omega^{4}\right)  .$ The extent of the $\left(  1/\sqrt{\omega}\right)$
 behavior will become larger as the range of the force
decreases.

In contrast to the spectrum of the position auto-correlation function, in the
case of a limited range force the higher frequency results are independent of
the damping.

\section{Conclusions}

We have found expressions for the spectrum of the correlation function of a pair of fields as seen by particles executing a 
continuous time, persistent random walk, with exponential distribution of
flight times in rectangular boundaries with specular reflection and single
velocity for 2 and 3 dimensions. The results are valid for all values of the
damping parameter $\lambda/L=\tau_{c}/\tau_{b},$ allowing calculations in the
ballistic and diffusive regimes as well as the transition region in between.

These conditional probability spectra allow the calculation of the spectrum of
the correlation functions of fields with arbitrary position dependence and
thus the study of relaxation and frequency shifts in arbitrary fields. The
results are valid as long as the trajectories are not influenced by the fields.

It is expected that the results will find applications in other fields such as
transport in mesoscopic systems.

\section{Acknowlegements}

R.G. and C.M.S. would like to thank Steve Clayton for his original work on
this problem and for helpful discussions. A.P. would like to thank Efim Katz
for helpful and stimulating discussions.

\end{document}